\documentstyle[epsf,subfigure]{mn}

\newif\ifAMStwofonts

\def\msun{{\rm\,M_\odot}}
\def\msun{{\rm\,M_\odot}}

\def\gyr{{\rm\,Gyr}}

\def\spose#1{\hbox to 0pt{#1\hss}}
\def\lta{\mathrel{\spose{\lower 3pt\hbox{$\mathchar"218$}}
     \raise 2.0pt\hbox{$\mathchar"13C$}}}
\def\gta{\mathrel{\spose{\lower 3pt\hbox{$\mathchar"218$}}
     \raise 2.0pt\hbox{$\mathchar"13E$}}}
\let\simless=\lta

\def\apj{\rm ApJ}

\def\apjl{\rm ApJL}
\def\aj{\rm AJ}
\def\mnras{\rm MNRAS}

\ifoldfss
  \ifCUPmtlplainloaded \else
    \NewTextAlphabet{textbfit} {cmbxti10} {}
    \NewTextAlphabet{textbfss} {cmssbx10} {}
    \NewMathAlphabet{mathbfit} {cmbxti10} {} 
    \NewMathAlphabet{mathbfss} {cmssbx10} {} 
  \fi
  \ifAMStwofonts
    \ifCUPmtlplainloaded \else
      \NewSymbolFont{upmath} {eurm10}
      \NewSymbolFont{AMSa} {msam10}
      \NewMathSymbol{\upi}     {0}{upmath}{19}
      \NewMathSymbol{\umu}     {0}{upmath}{16}
      \NewMathSymbol{\upartial}{0}{upmath}{40}
      \NewMathSymbol{\leqslant}{3}{AMSa}{36}
      \NewMathSymbol{\geqslant}{3}{AMSa}{3E}

       \let\le=\leqslant
       
    \fi
  \fi
\fi 

\ifnfssone
  \newmathalphabet{\mathit}
  \addtoversion{normal}{\mathit}{cmr}{m}{it}
  \addtoversion{bold}{\mathit}{cmr}{bx}{it}
  \newmathalphabet{\mathbfit} 
  \addtoversion{normal}{\mathbfit}{cmr}{bx}{it}
  \addtoversion{bold}{\mathbfit}{cmr}{bx}{it}
  \newmathalphabet{\mathbfss} 
  \addtoversion{normal}{\mathbfss}{cmss}{bx}{n}
  \addtoversion{bold}{\mathbfss}{cmss}{bx}{n}
  \ifAMStwofonts
    \ifCUPmtlplainloaded \else
      \UseAMStwoboldmath
      \makeatletter
      \new@mathgroup\upmath@group
      \define@mathgroup\mv@normal\upmath@group{eur}{m}{n}
      \define@mathgroup\mv@bold\upmath@group{eur}{b}{n}
      \edef\UPM{\hexnumber\upmath@group}
      \new@mathgroup\amsa@group
      \define@mathgroup\mv@normal\amsa@group{msa}{m}{n}
      \define@mathgroup\mv@bold\amsa@group{msa}{m}{n}
      \edef\AMSa{\hexnumber\amsa@group}
      \makeatother
      \mathchardef\upi="0\UPM19
      \mathchardef\umu="0\UPM16
      \mathchardef\upartial="0\UPM40
      \mathchardef\leqslant="3\AMSa36
      \mathchardef\geqslant="3\AMSa3E

       \let\le=\leqslant

    \fi
  \fi
\fi 

\ifnfsstwo
  \DeclareMathAlphabet{\mathbfit}{OT1}{cmr}{bx}{it}
  \SetMathAlphabet\mathbfit{bold}{OT1}{cmr}{bx}{it}
  \DeclareMathAlphabet{\mathbfss}{OT1}{cmss}{bx}{n}
  \SetMathAlphabet\mathbfss{bold}{OT1}{cmss}{bx}{n}
  \ifAMStwofonts
    \ifCUPmtlplainloaded \else
      \DeclareSymbolFont{UPM}{U}{eur}{m}{n}
      \SetSymbolFont{UPM}{bold}{U}{eur}{b}{n}
      \DeclareSymbolFont{AMSa}{U}{msa}{m}{n}
      \DeclareMathSymbol{\upi}{0}{UPM}{"19}
      \DeclareMathSymbol{\umu}{0}{UPM}{"16}
      \DeclareMathSymbol{\upartial}{0}{UPM}{"40}
      \DeclareMathSymbol{\leqslant}{3}{AMSa}{"36}
      \DeclareMathSymbol{\geqslant}{3}{AMSa}{"3E}

       \let\le=\leqslant

    \fi
  \fi
\fi 

\ifCUPmtlplainloaded \else
  \ifAMStwofonts \else 
    \def\upi{\pi}
    \def\umu{\mu}
    \def\upartial{\partial}
  \fi
\fi

\begin{document}

\title{Noise-driven evolution in stellar systems: \\ 
  A universal halo profile}

\author[Martin D. Weinberg]
{Martin D. Weinberg\\
  Department of Astronomy,
  University of Massachusetts, 
  Amherst, MA 01003-4525, USA}

\date{}
\pagerange{\pageref{firstpage}--\pageref{lastpage}}
\pubyear{}

\maketitle

\label{firstpage}

\begin{abstract}
  Using the theory describing the evolution of a galaxy halo due to
  stochastic fluctuations developed in the companion paper, we show
  that a halo quickly evolves toward the same self-similar profile,
  independent of its initial profile and concentration.  The
  self-similar part of profile takes the form of a double power law
  with inner and outer exponents taking the values near $-1.5$ and
  $-3$ respectively.  The precise value of the inner exponent depends
  on the magnitude and duration of the noisy epoch and most likely on
  form of the inner profile to start. The outer exponent is the result
  of evolution dominated by the external $l=1$ multipole resulting
  from the inner halo's response to noise.
  
  Three different noise processes are studied: (1) a bombardment by
  blobs of mass small compared to the halo mass (`shrapnel'); (2)
  orbital evolution of substructure by dynamical friction
  (`satellites'); and (3) noise caused by the orbit of blobs in the
  halo (`black holes').  The power spectra in the shrapnel and
  satellite cases is continuous and results in the double power law
  form, independent of initial conditions.  The power spectrum for
  black holes is discrete and has a different form with a much slower
  rate of evolution.  A generic prediction of this study is that noise
  from transient processes will drive evolution toward the same double
  power law with only weak constraints on the noise source and initial
  conditions.
\end{abstract}

\begin{keywords}
galaxies:evolution --- galaxies: haloes --- galaxies: kinematics and
dynamics --- cosmology: theory --- dark matter
\end{keywords}

\section{Introduction}
\label{sec:intro}

There is considerable evidence from numerical experiments that haloes
formed in CDM simulations are well-approximated by two-component power
law profiles of the form $\rho\propto r^{-\gamma}(1+r/r_s)^{\gamma-3}$
or $\rho\propto r^{-\gamma}(1+(r/r_s)^{3-\gamma})^{-1}$.  The first
form was presented by Navarro, Frenk \& White (1997, hereafter NFW)
based on a suite of simulations with different initial density
fluctuation spectra and cosmological parameters.  They suggest
$\gamma=1$ is the universal exponent. Moore et al.  (1998) find that
the form of the inner profile ($\gamma$) is resolution dependent.
With $3\times10^{6}$ particles within the virial radius, this group
finds $\gamma=1.4$.  Jing \& Suto (2000) find that the inner slope is
not universal but varies depending on environment; they find
$\gamma=1.5, 1.3$ and $1.1$ for galaxy-, group-, and cluster-mass
haloes, respectively.  Besides issues of n-body resolution and
methodology, attempts to explain the discrepancy of these
collisionless simulations and astronomical observations include new
laws of physics (Spergel \& Steinhardt 2000) and the effects of gas
dissipation (Tittley \& Couchman 1999, Frenk et al. 2000, Alvares,
Shapiro \& Martel 2000).

The suggestion of some sort of universal profile and more generally
the physics of dissipationless collapse or violent relaxation has a
long history.  The general problem of stellar dynamics in the presence
of large fluctuations is very difficult and much of this focuses on
first-principle forms for the phase-space distribution.  The companion
paper approaches this problem as near-equilibrium evolution in a noisy
environment (see Weinberg 2000, Paper 1, for a review of recent
theoretical work on this subject).  Here we apply the theory from
Paper 1 to follow the evolution of haloes with a number of different
concentrations and shapes.  The near-equilibrium restriction allows
development of a solvable evolutionary equation given some initial
condition.  We find that the profile rapidly assumes a double power
law form with $\gamma\approx1.5$ independent of the initial model.  So
in the presence of noise, a quasi-self-similar profile does appear and
in this sense represents the near-equilibrium limit of violent
relaxation.  The inner power-law profile evolves slowly if the noise
is applied over long periods.  This evolution of the inner halo may
depend on the details of the central initial conditions (e.g.
power-law cusp or core) and possibly the noise source.  Investigation
of this issue is underway.

The plan for this paper is as follows.  The basic principles from
Paper 1 are described and reviewed in \S\ref{sec:physics} followed by
a description of the noise models in \S\ref{sec:models}.  Although we
choose several particular astronomical scenarios to derive particular
noise spectra, the resulting halo profiles are insensitive to the
shape of the power spectrum for noise due to transient perturbations;
the dynamics behind this finding is described.  We also consider an
example of non-transient noise source, a halo of massive black holes.
This case does not result in the double power law form and leads to
much weaker evolution overall.  The resulting halo profiles are
described in \S\ref{sec:results} followed by a discussion in
\S\ref{sec:discussion} and summary in \S\ref{sec:summary}.

\section{Physics of noise-driven evolution}
\label{sec:physics}

Let us begin by considering a perturbation to a galaxy halo.  For
example, a dwarf passing through the halo. The halo responds with a
wake whose form depends on the halo profile, perturber velocity and
minimum impact parameter (the dynamics for a single encounter has been
considered in detail by Vesperini \& Weinberg 2000).  The
gravitational attraction of the halo wake and the interloping dwarf
exchanges energy and angular momentum between the two.  This new
momentum is deposited (or removed) in the dark matter halo at the
location of the wake.  After the dwarf is gone, the halo will adjust
its profile to reachieve equilibrium\footnote{For encounters in the
  outer halo, note that the phase mixing time may be longer than the
  age of the Universe but this will not affect the current study.}.
As long as the wake is relatively small, many such transient
encounters may be in progress simultaneously without mutual
interaction.  We may now ask the question: can we compute the mean
evolution after many such encounters without resorting to simulating
an ensemble of encounters at high resolution?

The answer to this question motivates the development of an evolution
equation that may be solved numerically but without simulation (see
Paper 1).  The resulting evolution equation takes the Fokker-Planck
form, similar to the kinetic equation used in studying globular
cluster evolution, although the details are cumbersome and the
approach somewhat different.  The shape of the wake (or ensemble of
wakes for different types of encounters) will determine the diffusion
coefficients and subsequently the shape of the evolving halo profile.

This shape of the wake is key to understanding how noise can drive a
self-similar evolution.  Both Weinberg (1998) and Vesperini \&
Weinberg (2000) illustrate an important fact in the dynamics of hot
collisionless systems: a halo has a low-order $m=1$ and to a lesser
extent $m=2$ weakly damped mode.  Because these modes damp slowly,
they tend to dominate the wake.  Weinberg (1998 and subsequent work)
shows that the scale of $m=1$ mode is proportional to the
characteristic radius of a halo profile.  The interaction between the
wake and the disturbance changes the actions of the orbits in the
modal excitation.  In the case of the fly-by example above, the energy
of these orbits tend increase over the range of the mode and this
modifies the halo profile.  The new characteristic radius then governs
the size and shape of subsequent modal excitations.  The direct
calculation shows that this process rapidly become self-similar and
results in a new power law slowly working its way out in the halo.
Although it was not obvious a priori that this would be the result of
the noise-driven evolution, we will see that it describes the
numerical solutions in \S\ref{sec:results}.

We consider two general types of noise perturbations in this paper:
transient and quasi-periodic perturbations.  These cover the types of
astronomical scenarios that are likely to be important.  In addition
to the dwarf fly-by in the example above and orbital decay which will
be considered explicitly in this paper, a {\em transient} event might
be a damping disk density wave or bar instability.  Astronomically
relevant {\em quasi-periodic} perturbations result from something
orbiting in the galaxy halo and such a halo of super massive black
holes whose individual masses are too small for dynamical friction to
be significant.

For a transient perturbation, the relative distribution of power over
all frequencies will depend on its details.  However, the continuous
spectra perturbations will nearly always have some power near the
frequencies of weakly damped modes.  In the halo case, the response is
dominated by the weakly damped $m=1$ mode by at least an order of
magnitude.  Therefore, no matter what shape the frequency spectrum
takes, the response will be similar.  This, together with description
of self-similar evolution above, is at the root of the surprising
claim that noise can result in a nearly universal profile, independent
of its source.
    
\begin{figure*}
  \centering
  \subfigure[Close to resonance]{\epsfxsize=3.25in\epsfbox{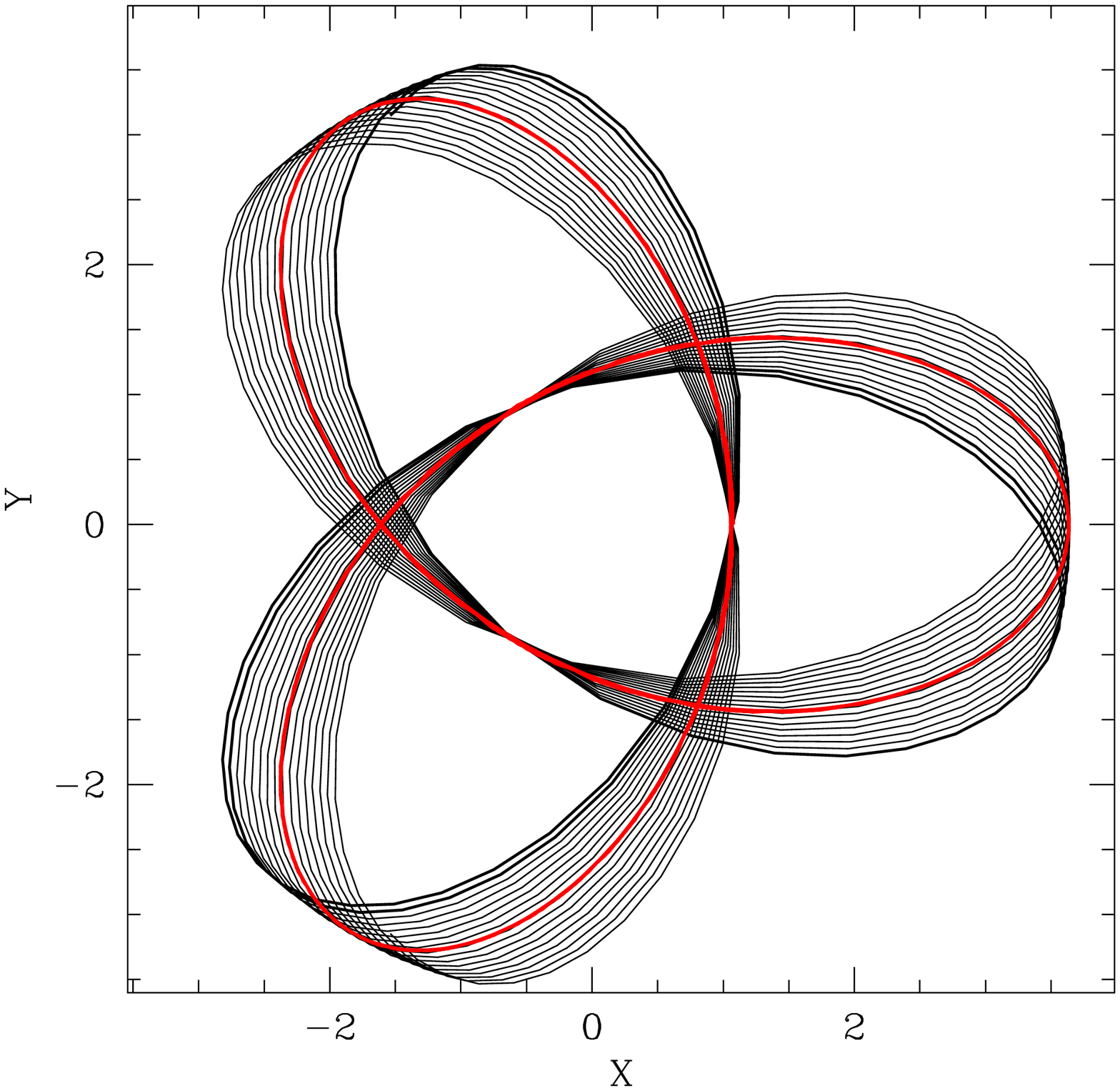}}
  \subfigure[Off resonance]{\epsfxsize=3.25in\epsfbox{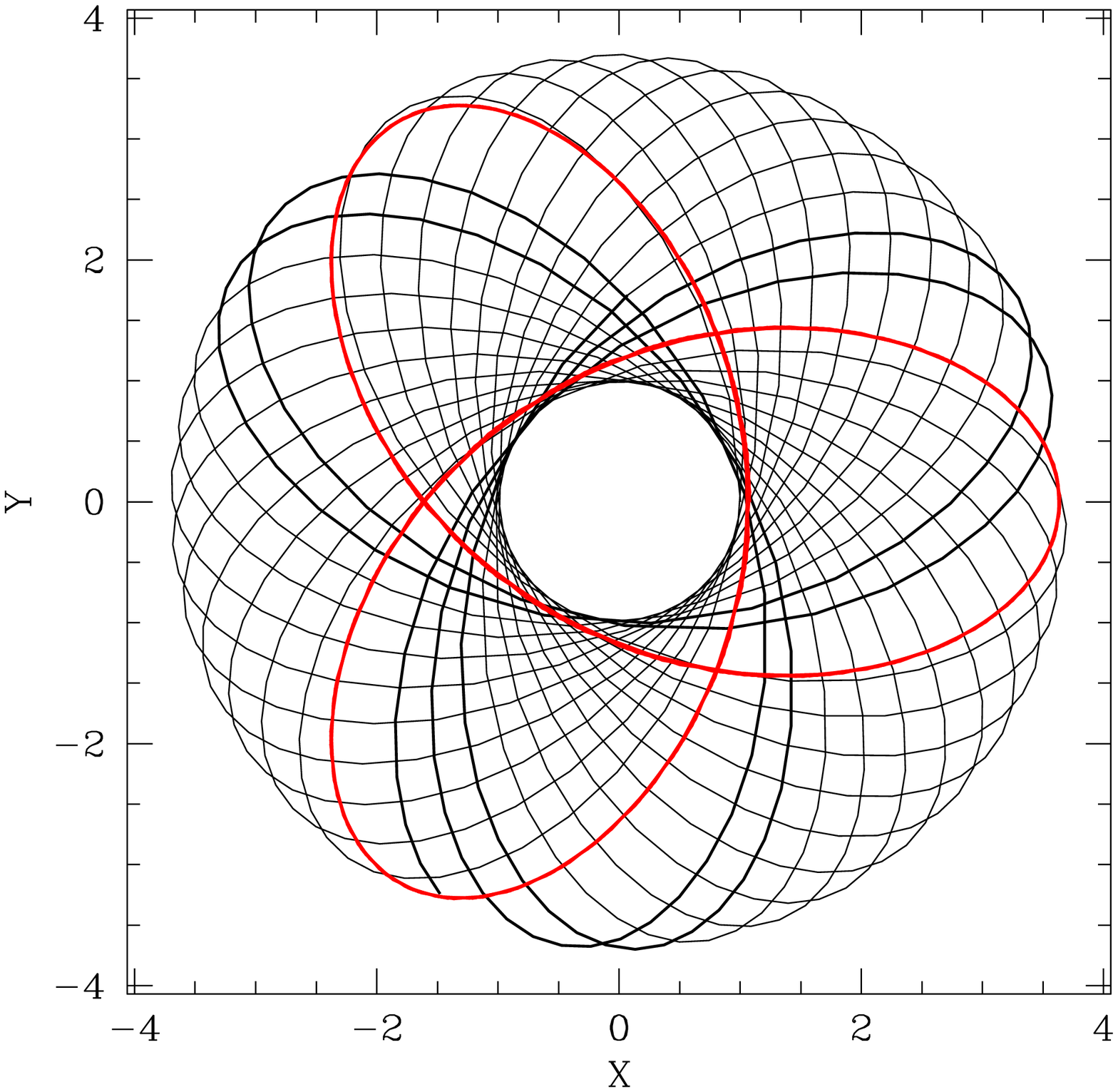}}
  \caption{Orbit close to zero-frequency low-order resonance ($n=-3,
    m=2$): (a) very close to resonance; (b) slightly off resonance.
    The heavy curve marks the closed resonant orbit.
    \label{fig:res23}}
\end{figure*}

For a quasi-periodic perturbation, the power will be at discrete
frequencies.  Each dark matter orbit only changes if driving
frequencies couple to its natural orbital frequencies.  For orbiting
point masses, these resonances must have zero frequency in order to
apply a torque; in other words, the resonant orbits are closed orbits.
If the perturber orbit is not closed, then over a long period of time,
its density will be axisymmetric and can not result in angular
momentum and energy exchange (see Fig.  \ref{fig:res23} for an
pictorial example).  For a physically plausible halo, there will be no
closed orbits at orders $l=1$ and $2$ for the following reasons.  Our
closed orbit condition is $n\Omega_1+m\Omega_2=0$ where $\Omega_1$ and
$\Omega_2$ are the radial and azimuthal orbital frequencies.  The
ratio $\Omega_1/\Omega_2$ ranges from 1 to 2 in most profiles.  For
$l=1$, we have $|m|\le1$ and a closed orbit for $m=1$ will only then
obtain for a point mass potential.  Similarly, for $l=2$, $|m|\le2$
and a closed orbit for $m=2$ will only then obtain at the centre of a
homogeneous core.  Both of these conditions are represented by a
vanishingly small number of orbits and therefore there are no resonant
contributions by internally orbiting point masses for harmonics with
$l<3$.
    
\section{Models and noise sources}
\label{sec:models}

We consider three classes of initial profiles: King models (1966) with
concentrations $c$ between 0.67 and 1.5, Plummer models (e.g.  Binney
\& Tremaine 1987), and models of the form:
\begin{equation}
  \rho \propto {1\over (r+\epsilon)^\gamma}{1\over (r+1)^{\beta-\gamma}}.
  \label{eq:NFWlike}
\end{equation}
Solution of the evolution equation requires repetitive integration of
the distribution function to derive the density, mass and potential
profiles.  For this reason, the dynamic range of cuspy models present
considerable numerical challenge and are not considered here,
unfortunately.  However, the instantaneous fluctuation spectrum in
models with and without cores exhibit strong enhancement due to $l=1$
modes (see Weinberg 1998); this suggests that qualitative behavior
here will be similar as well since the underlying physical process is
the same.  The model described in equation (\ref{eq:NFWlike}) has most
of the features of popular cusp models\footnote{E.g. equation
  (\ref{eq:NFWlike}) with $\epsilon\rightarrow0$, $\gamma=1$ and
  $\beta=3$ is the NFW profile} but gets around numerical difficulties
of divergent distribution phase-space distribution functions for
carefully chosen values of small but non-vanishing values of
$\epsilon$.  It remains a possibility that an initial cusp may effect
the sense of the evolution of the inner power law.  We will extend the
development to study this case in detail in a later paper.  Here, we
focus on the rapid approach to a self-similar form ouside of the core.

We consider three specific noise sources: (1) transient noise due to
blobs moving on rectilinear trajectory (`shrapnel' model); (2)
transient noise due to substructure on decaying halo orbits due to
dynamical friction (`satellite' model); and (3) quasi-periodic noise
due to blobs orbiting within the halo (`black hole' model).  The
derivation of moments for the evolution equation can be found in \S4
of Paper 1.  For all noise sources, the amplitude for a single event
is proportional to the square of the mass in the perturbation since
the net change in conserved quantities is second order.  From the form
of the collisional Boltzmann equation, the evolutionary time scale is
inversely proportional to amplitude of the collision term.  This
allows easy scaling of the results derived here for each noise source
described below.  We will set some fiducial parameters for results
quoted in \S\ref{sec:results} and give the scaling formula for the
amplitudes so that the time scales can be easily derived for other
scenarios.

For the shrapnel model, the overall amplitude for the process is also
proportional to the bombardment rate.  We assume that the flux of
shrapnel is uniform so that the distribution impact parameters $b$ is
proportional to $b$.  Soon after formation, one expects encounters to
be more numerous so we adopt a fiducial rate of 10 encounters per
gigayear inside of 50 kpc, each with a mass of 0.003 halo masses.
Scaling to our Galaxy, our fiducial shrapnel has 15\% of the LMC mass.
The trajectories have constant velocity chosen to be $\sqrt{2}$ times
the peak halo circular velocity.  However, the results are nearly
unchanged if the incoming velocity is increased or decreased by a
factor of two and so this is not a sensitive assumption.

For the satellite model, we assume a halo's worth of satellites of a
given mass assimilating within 1 Gyr.  Because the amplitude is
proportional to the square of the satellite to halo mass ratio but the
number of satellites is proportional to the inverse of this ratio, the
overall amplitudes scales as the satellite to halo mass ratio.  This
scaling is roughly consistent with the substructure distribution
described by Moore et al. (1999).  For smaller numbers of satellite
per halo, the overall amplitude can be multiplied by the desired
factor which lengthens the time scale proportionately.  The noise
spectrum results from following the orbital decay of a satellite of
given mass ratio by direct integration of the equations of motion of
an initially circular orbit at a radius enclosing 95\% of the halo
mass.  The drag force is computed using Chandrasekhar's formula with
$\ln\Lambda=8$; this value provides a good match to substructure
simulation (Tormen et al. 1998).  Although the power spectrum is
computed for the full orbital decay of the satellite, the amplitude is
diminished by the fraction of satellites with orbits that can fully
decay in one gigayear. Mass loss from the satellite is not included.

In the case of the black hole model, the amplitude of the noise also
is proportional to the square of the perturber mass for a single
perturber.  However, the amplitude is also proportional to number of
perturbers.  For a fixed fraction in black holes, the number is then
inversely proportional to the black hole mass.  Altogether, then, the
amplitude is directly proportional to the perturber mass and the
fraction of the halo represented by the perturber mass.  For our
fiducial example, we assume a halo fully populated by $10^6$
solar-mass black holes (Lacey \& Ostriker 1985).

The scaling and fiducial parameters for all of these cases is
summarized in Table \ref{tab:scale}.

\begin{table*}
\centering
\caption{Scaling and fiducial parameters}
\label{tab:scale}
\begin{tabular}{lll}
\\ \hline
\multicolumn{3}{l}{Shrapnel model: $A=A_o
  (R_{enc}/R_{enc\,0})(m_{enc}/m_{enc\,0})^2$} \\ \hline
  $R_{enc}$ & Number of encounters within 50 kpc per Gyr & 10 \\
$m_{enc}$ & Relative mass of shrapnel (units of halo mass) & 0.03 \\ \hline
\multicolumn{3}{l}{Satellite model: $A=A_o N_{Halo}
  (m_{sat}/m_{halo})^2 (\ln\Lambda/8.0)^2$} \\ \hline
$m_{sat}/m_{halo}$ & Satellite to halo mass ratio & 0.01, 0.03, 0.05 \\
  $N_{halo}$ & Number of satellites accreting per Gyr &
  $m_{halo}/m_{sat}$ \\  \hline
\multicolumn{3}{l}{Black hole model: $A=A_o
  (f_{bh}/f_{bh\,0})(n_{bh}/n_{bh\,0})$} \\ \hline
$n_{bh}$ & Relative mass of body (units of halo mass) & $10^6$ \\
$f_{bh}$ & Fraction of halo in lumps & $1.0$ \\ \hline
\end{tabular}
\end{table*}

\section{Resulting halo profiles}
\label{sec:results}

The evolution equation, a Boltzmann equation with a Fokker-Planck-type
collision term, is derived in Paper 1.  To simplify the numerical
solution, the full equation is isotropized by averaging over angular
momentum to leave an equation for the phase-space distribution
function in energy $E$ and time $t$ (see Appendix \S\ref{sec:FP} and
Paper 1).  The Fokker-Planck equation is solved on a grid as described
in Appendix \S\ref{sec:numerical}.  The energy variable $E$ is
remapped to better populate the centre and outer halo with grid
points.  This new mapping variable $x=x(E)$ is monotonic in energy and
described in Appendix \S\ref{sec:mapping}.

Figure \ref{fig:densevol} shows the evolution of four different
initial models under shrapnel noise.  The radial scaling is chosen to
place the initial half-mass radius at approximately 50 kpc.  The main
features are as follows:
\begin{enumerate}
\item In all cases, there are two distinct evolutionary phases: (1) a
  transient readjustment to a double power law profile; (2) slow,
  approximately self-similarly evolution of the double power law
  profile.  The transition period will be described in more detail
  below.
\item The asymptotic outer power law exponent is -3.  The profile
  continues to approach the -3 form at increasing radius as the
  evolution continues.
\item The inner power law exponent is approximately -1.5 after the
  transient readjustment phase and slowly decreases thereafter.  For
  example, after 1.9 Gyr in Panel (a) in Figure \ref{fig:densevol},
  the inner power law is roughly -1.2.  The exponent has a value near
  -1.5 for the initially steeper profile shown in Panel (c).
\item The more concentrated models, which have deeper potential wells
  and therefore shorter dynamical times, evolve most quickly.  The
  softened double power law model (eq.  \ref{eq:NFWlike}) has a
  considerably shorter evolutionary time scale (see below) then any of
  the King or Plummer profiles.
\end{enumerate}

These findings together with an examination of the contribution of
specific harmonics offer insight into the origin of the profile.  The
input energy from the fly by or orbital decay moves mass outwards,
expanding the profile.  The $r^{-3}$ profile occurs outside of the
modal peak in the asymptotic outer power law tail of the $l=1$
multipole.  Because all initial conditions and transient noise sources
result in the $r^{-3}$ form, we conclude that it is the self-similar
response of the halo to the outer $l=1$ multipole.  Higher order
multipoles ($2\le l\le6$ were checked explicitly) do not result in the
$r^{-3}$ profile and have much smaller relative amplitudes.  This
suggests an explanation for the ubiquity of the $r^{-3}$ outer
profile: many noise source will excite $l=m=1$ modes which cause the
profile and since it will be independent of the noise source or
initial profile. The coincidence of this profile with those found in
cosmological n-body simulations further suggests that $l=1$ noise
plays a significant role in halo formation.

The approximate inner exponent of $-1.5$ appears near the peak of the
mode and is not simply analyzed.  The shape of the mode depends
instantaneously on the profile.  The mode then absorbs energy from the
noise source and transports mass outward which in turn shapes the
mode.  Repetitive excitation leads to a self-similar profile near the
peak of mode.

Because these models have cores, and both the radial and azimuthal
orbital frequencies are nearly the same in the core, it is difficult
to couple to these orbits in order to transfer angular momentum in and
out of the core.  The core, then, expands with the overall expansion
of the halo due to the deposition of energy from the noise sources.
These dynamics suggest that we restrict our consideration to evolution
beyond the core, as described earlier.

Figure \ref{fig:massevol} describes the distribution of mass in the
mapped energy grid $X$.  The quantity $F(X)$ is the phase-space
distribution function in the mapped variable $X$ and $J(X)$ describes
the phase-space volume between $X$ and $X+dX$.  Therefore $F(X)\times
J(X)$ is the differential mass.  As the potential well evolves, the
mapping is rescaled so that inner point has the same value.  The first
panel in Figure \ref{fig:massevol} shows the evolution of the mass
distribution as it attains the double power law form.  For this
initial condition, mass is shift outward to populate the higher energy
tail.  During the second self-similar phase, the double power law
profile is established (note the constancy of the profile between
$X=-2$ and $X=0$) and further evolution populates the outer tail,
driving the $r^{-3}$ profile to larger radii.

\begin{figure*}
  \centering
  \subfigure[King $W_0=3$]{\epsfxsize=2.8in\epsfbox{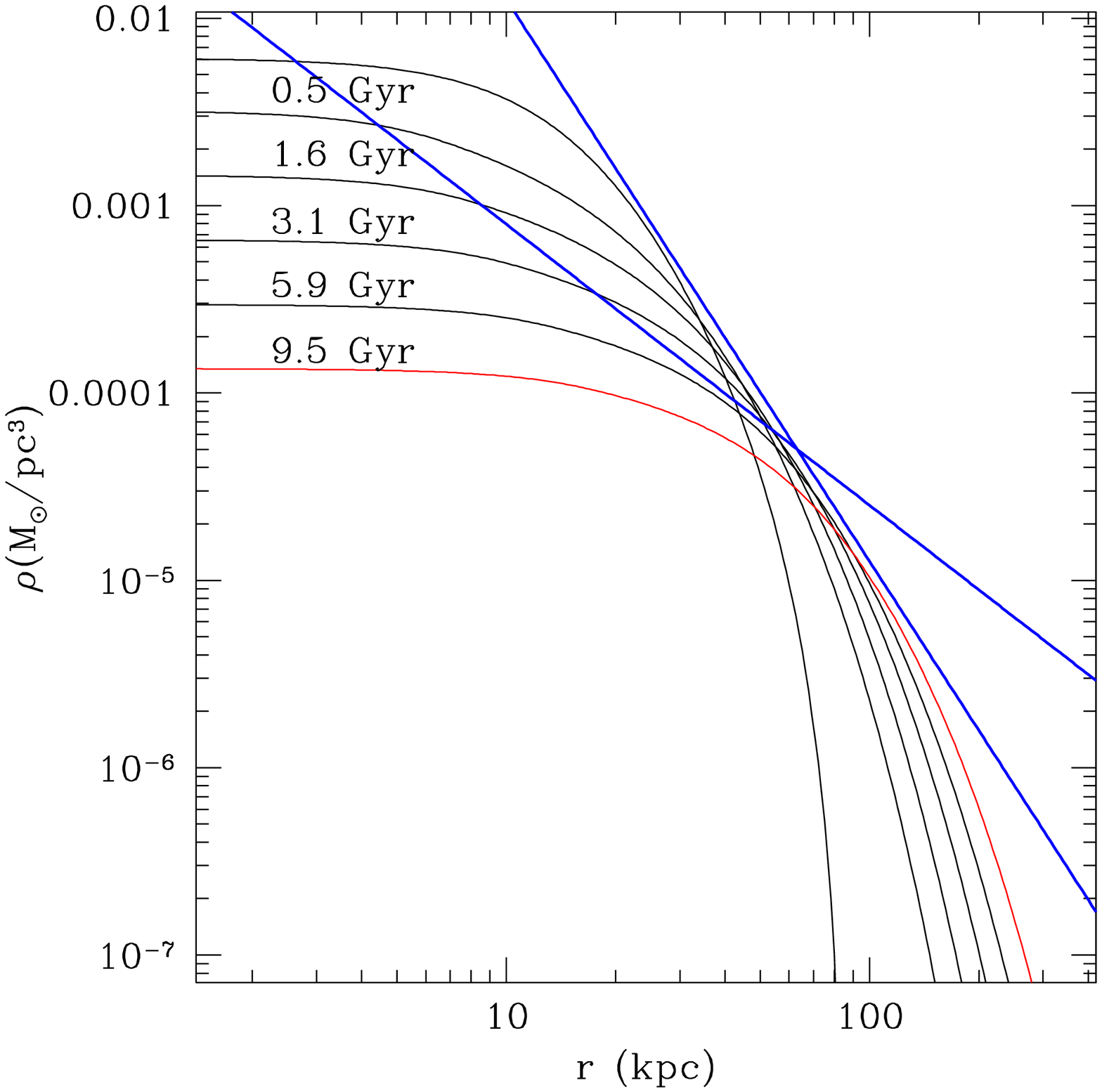}}
  \subfigure[King $W_0=5$]{\epsfxsize=2.8in\epsfbox{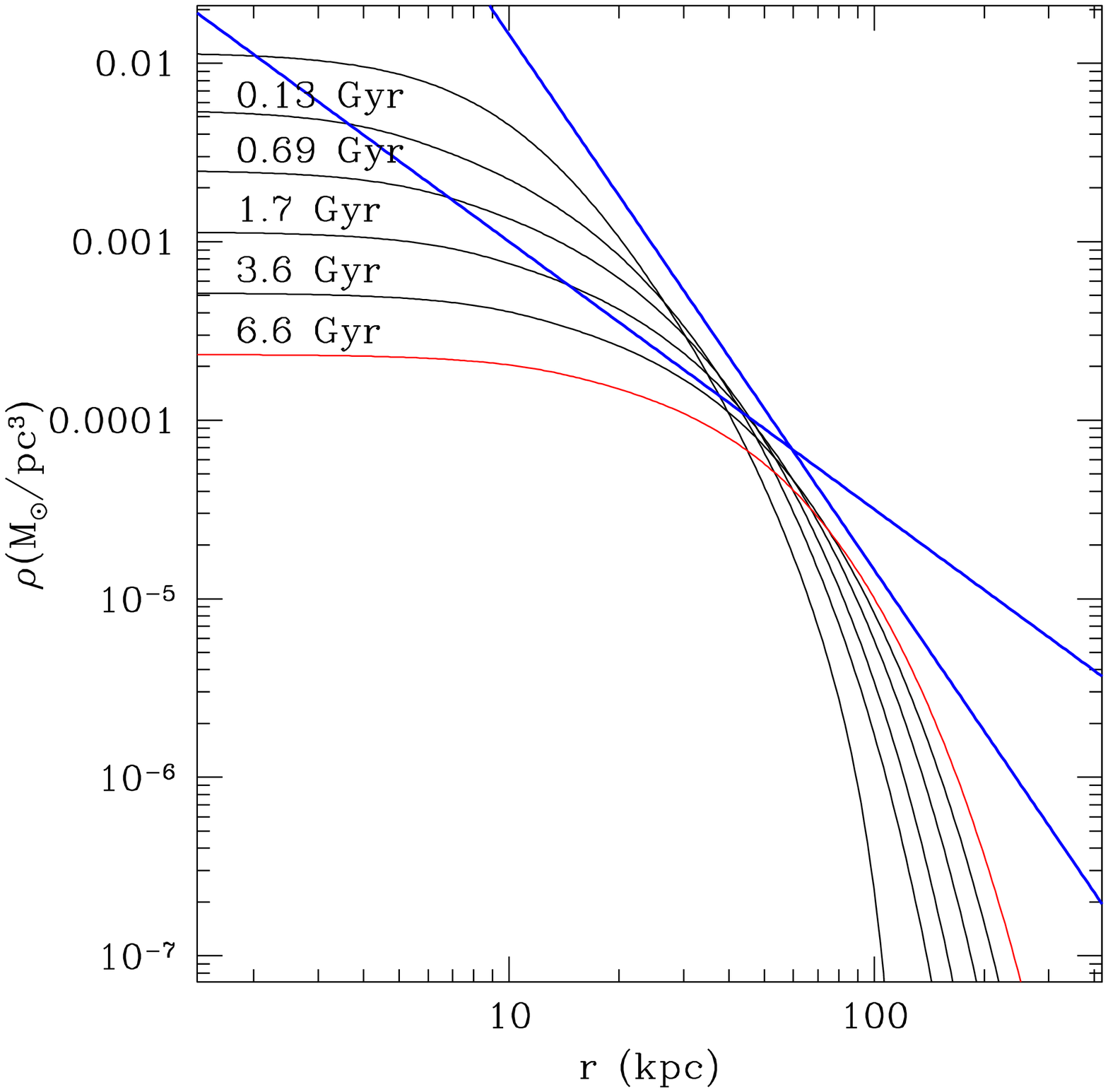}}
  \subfigure[King $W_0=7$]{\epsfxsize=2.8in\epsfbox{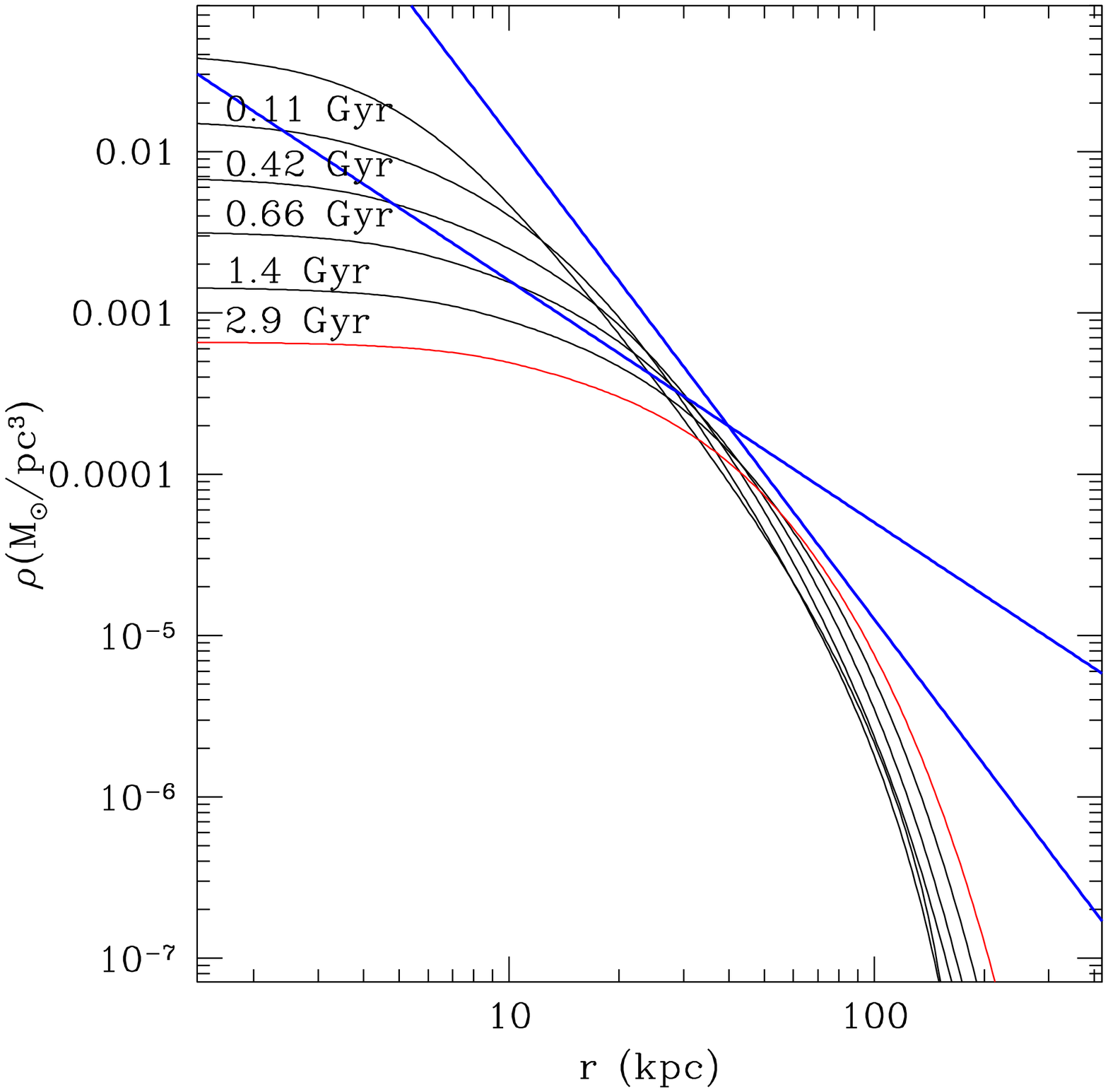}}
  \subfigure[Plummer]{\epsfxsize=2.8in\epsfbox{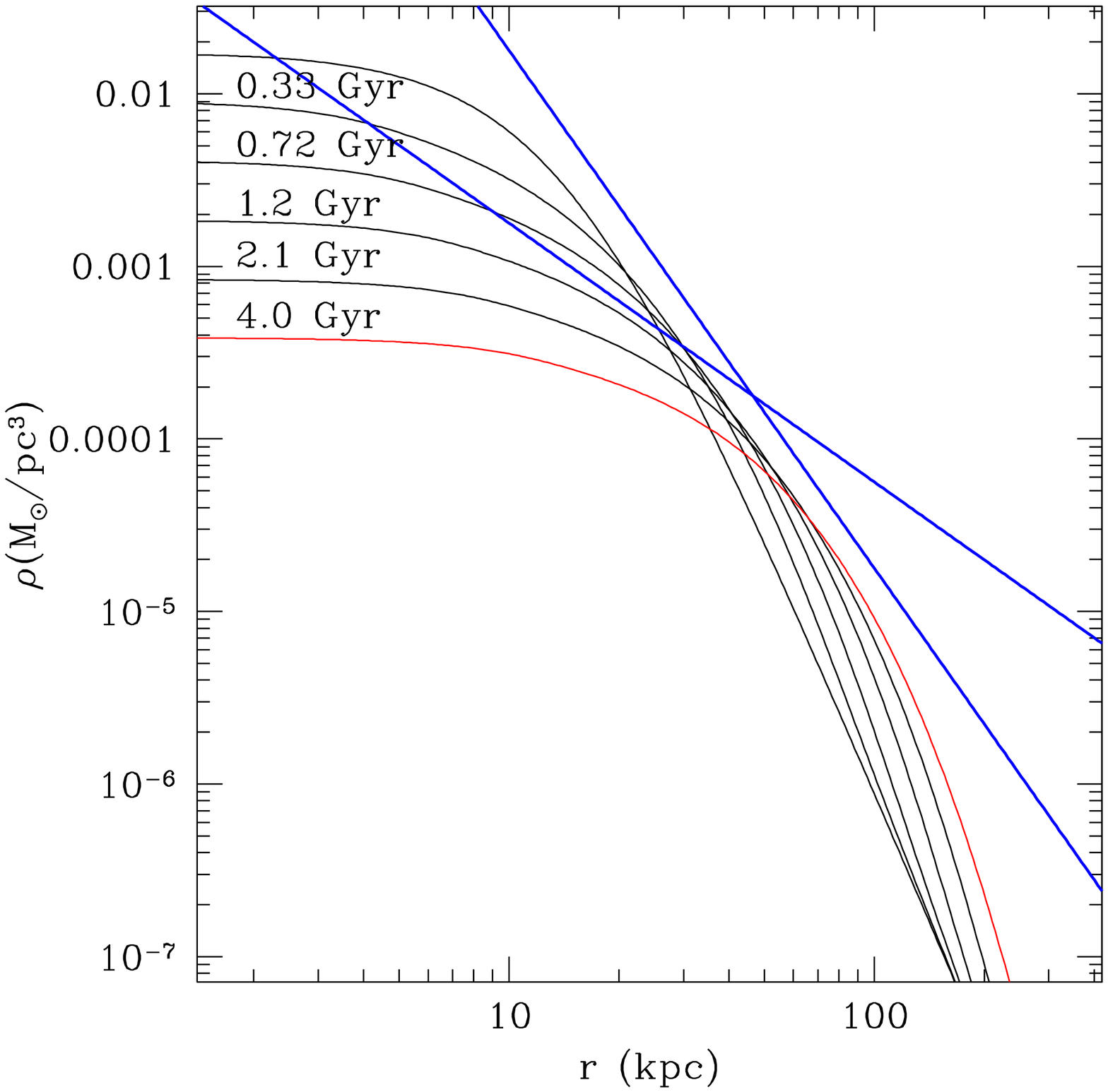}}
  \caption{Evolution of the initial profile (labelled) under
    `shrapnel' noise.  The top curve in each case is the initial
    profile. The times for each curve are derived using the fiducial
    scaling from Table \protect{\ref{tab:scale}}.  The heavy straight
    lines show power laws with exponent -1.5 and -3.0 for comparison.}
  \label{fig:densevol}
\end{figure*}

\begin{figure*}
  \centering
  \subfigure[Approach to power law profiles]{\epsfxsize=3.0in\epsfbox{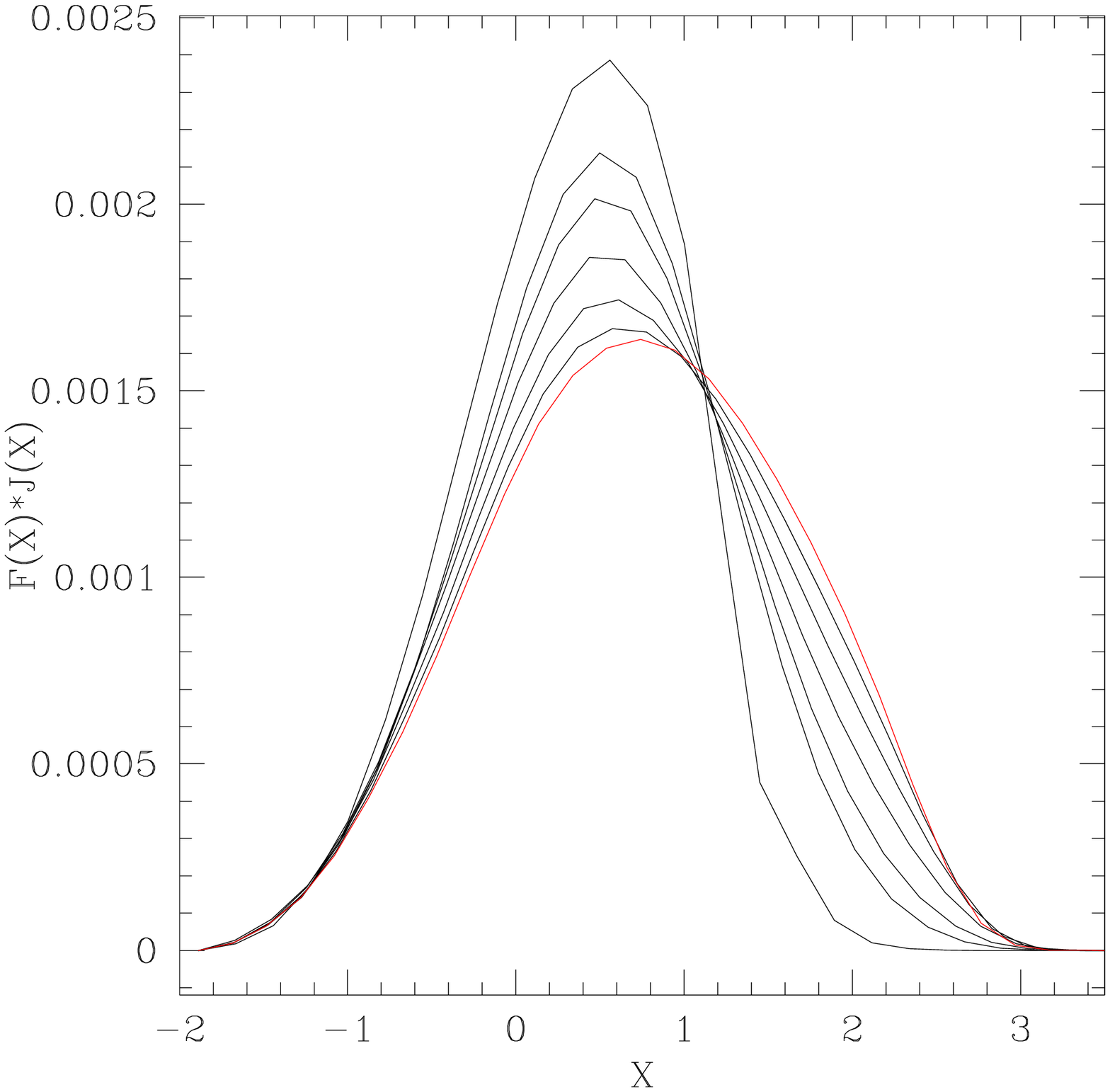}}
  \subfigure[Self-similar phase]{\epsfxsize=3.0in\epsfbox{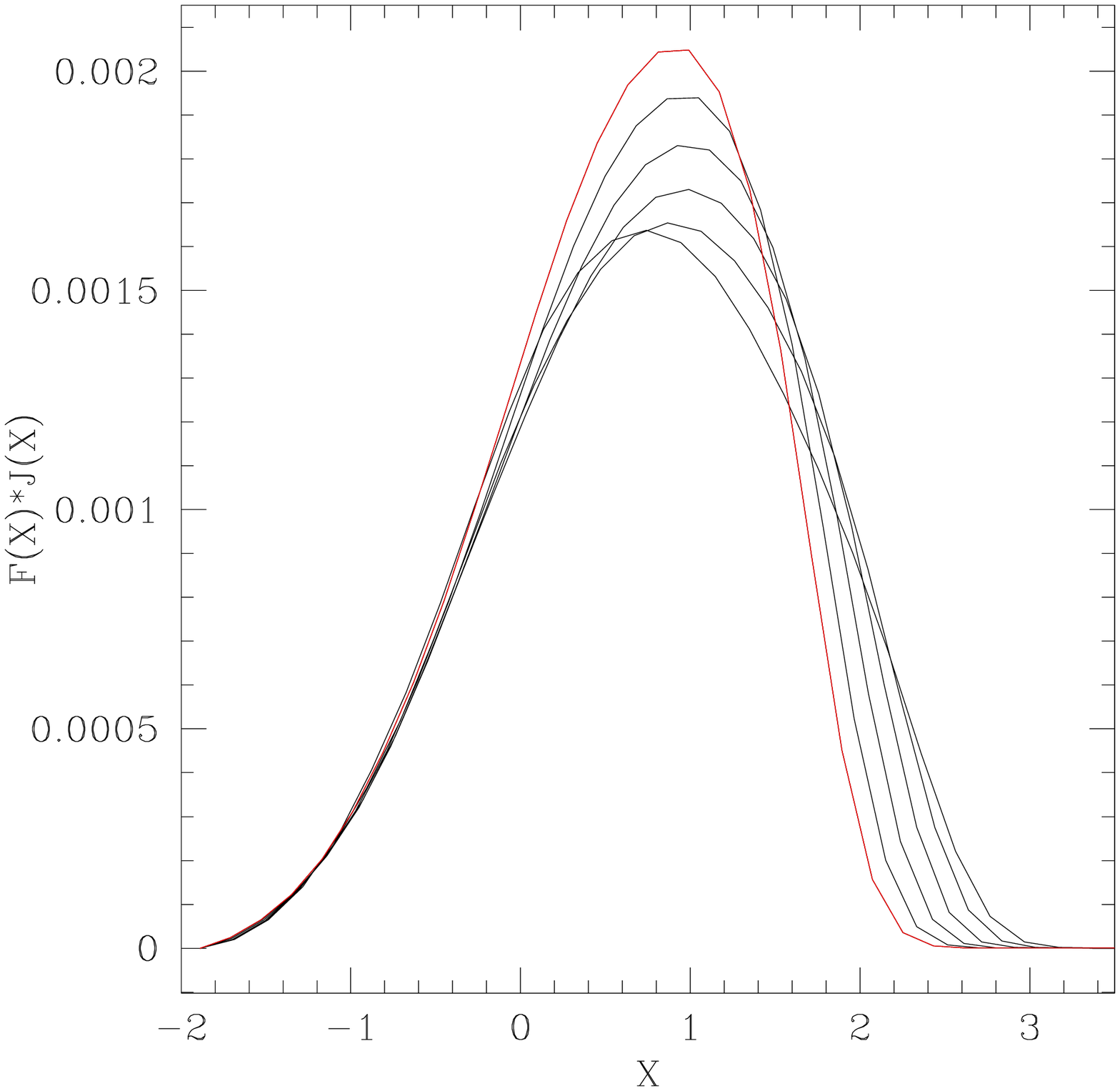}}
  \caption{For each initial condition
    (cf. Fig. \protect{\ref{fig:densevol}}), there are two phases of
    evolution.  In the first phase, the model approaches the double
    power law profile; the model spreads out in $x$.  In the second
    phase, the double power law slowly evolves; the inner mass
    distribution evolves self-similarly in $x$ with mass moving
    outward while the peak at $x\approx1$ grows.  Panels (a) and (b)
    show the two phases separately.  For each of the cases in Fig.
    \protect{\ref{fig:densevol}}, the approach phase is completed by
    1.0, 0.5, and 0.9 gigayear, respectively.}
    \label{fig:massevol}
\end{figure*}

\begin{figure*}
  \centering
  \subfigure[$m_{sat}/m_{halo}=0.05$]{\epsfxsize=2.25in\epsfbox{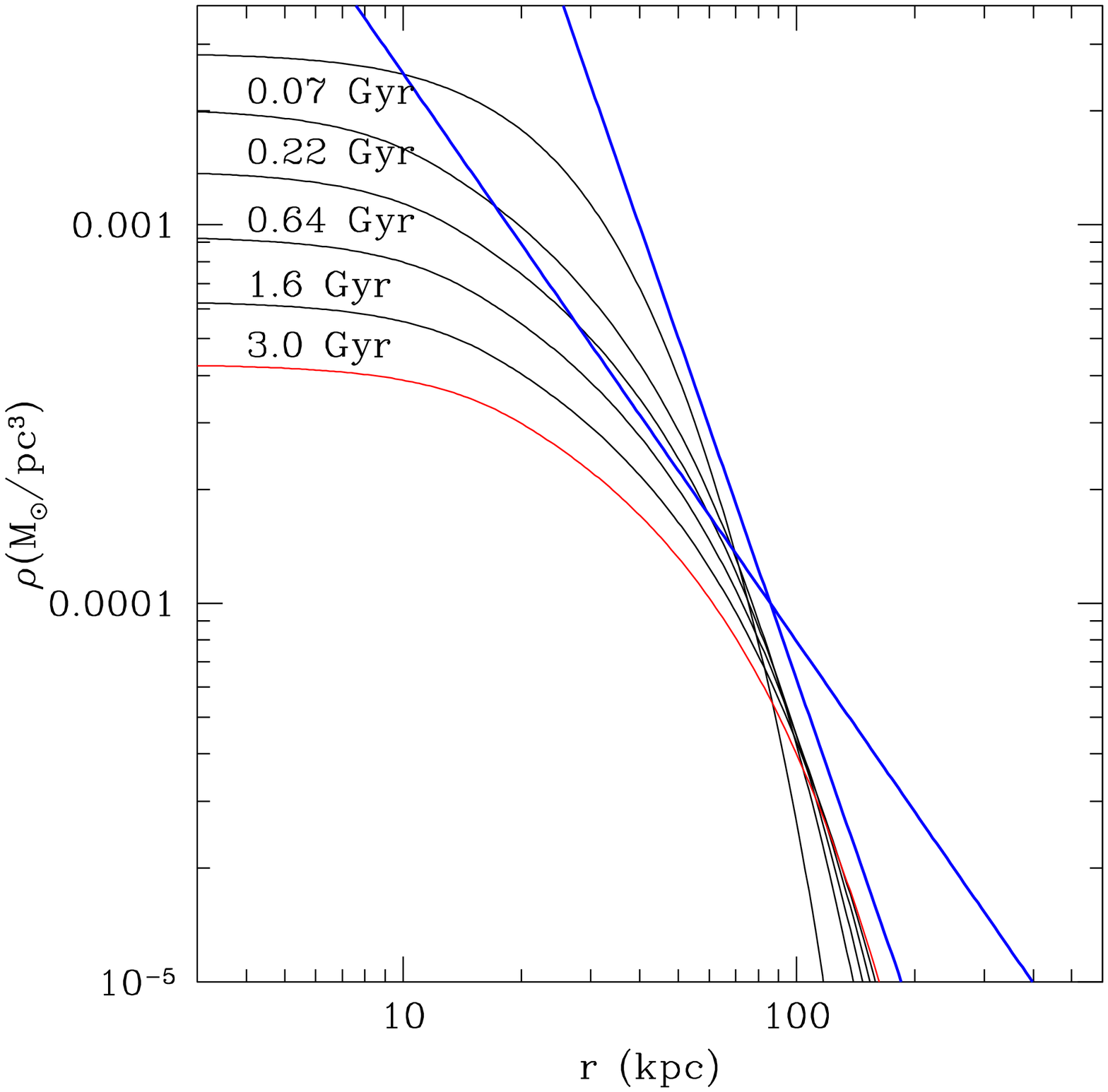}}
  \subfigure[$m_{sat}/m_{halo}=0.03$]{\epsfxsize=2.25in\epsfbox{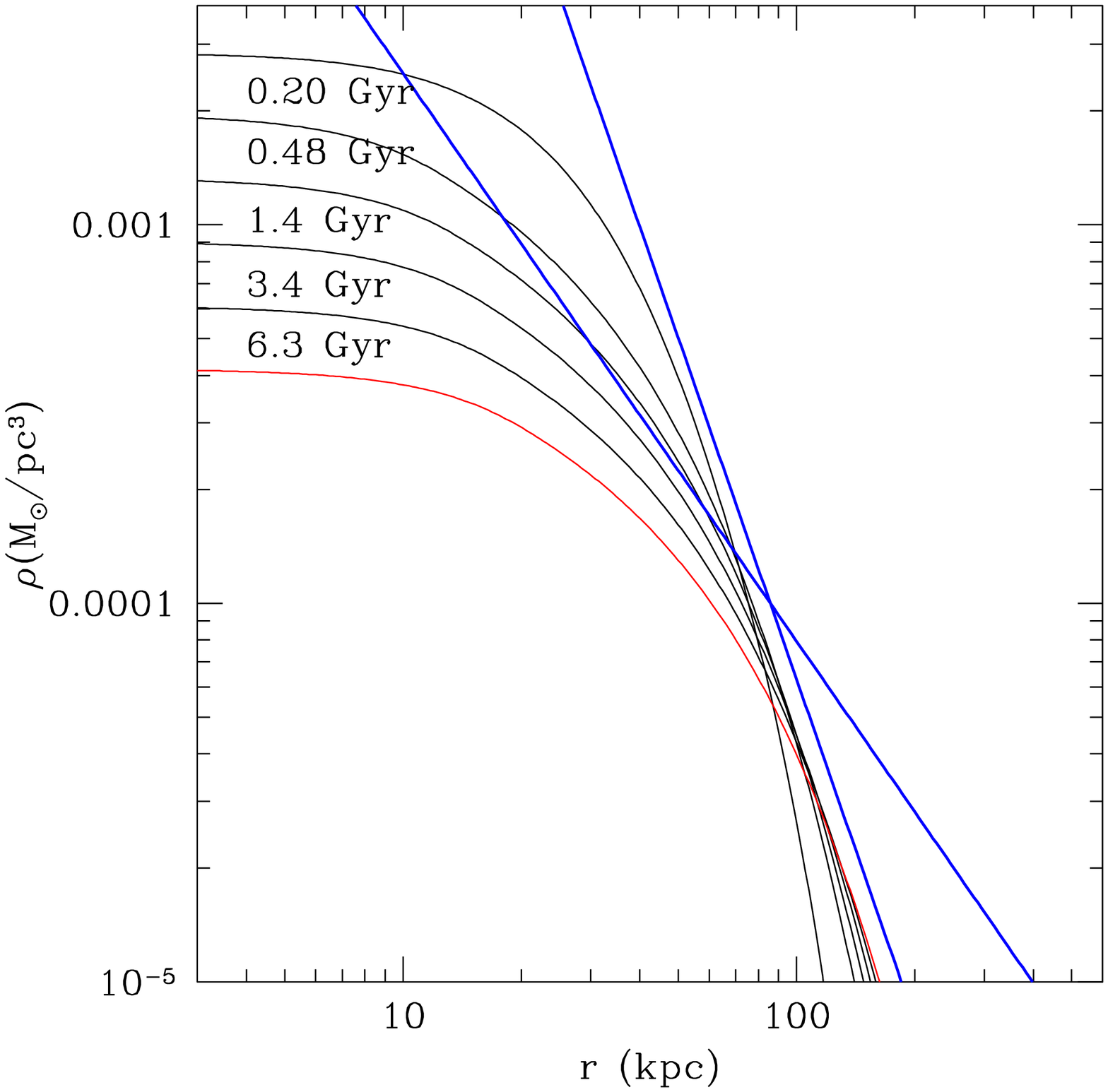}}
  \subfigure[$m_{sat}/m_{halo}=0.01$]{\epsfxsize=2.25in\epsfbox{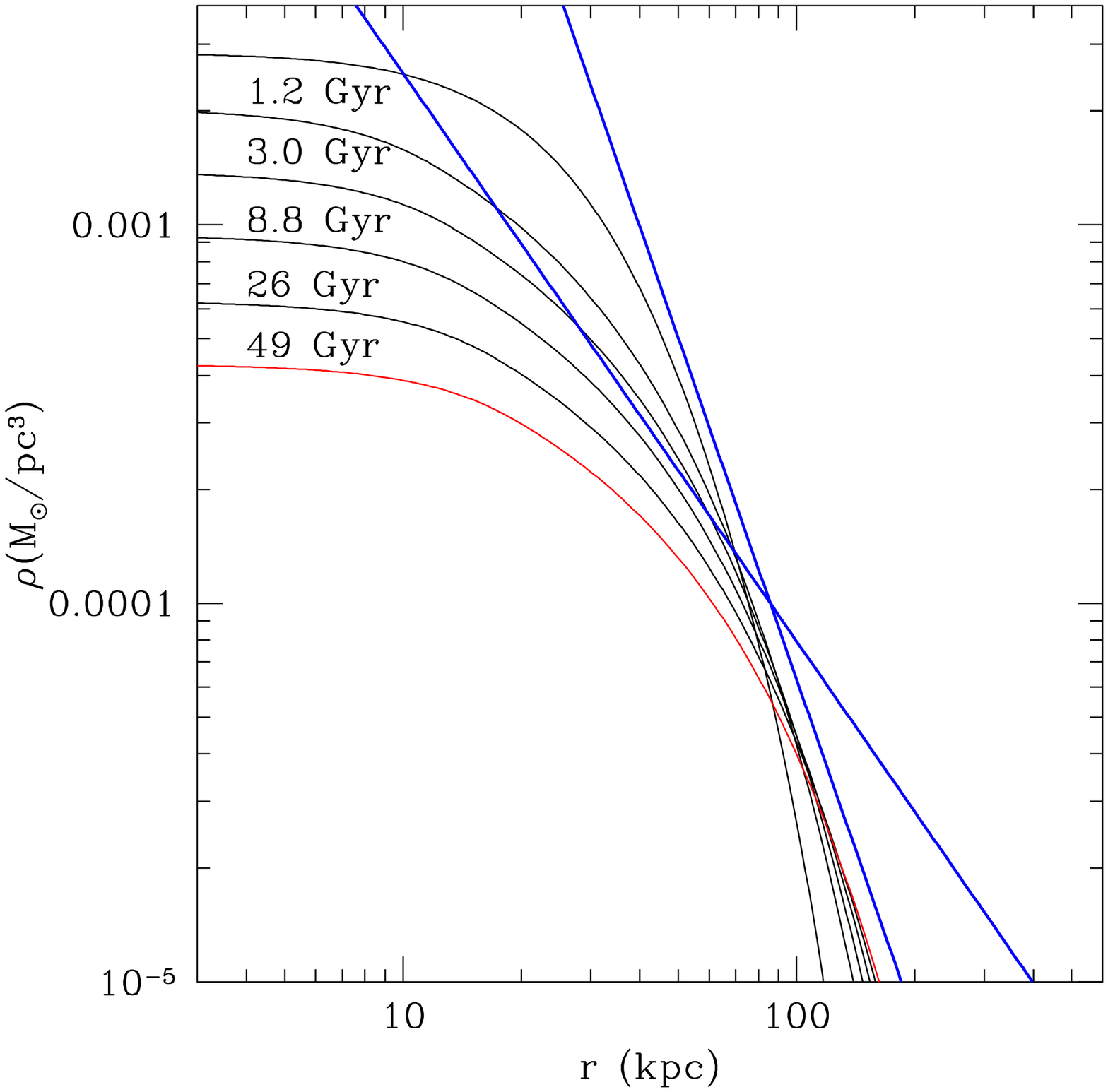}}
  \caption{As in Figure \protect{\ref{fig:densevol}}) but for noise
    due to decaying satellite perturbers using the fiducial
    parameterization from Table \protect{\ref{tab:scale}}.  Panels (a),
    (b) and (c) show satellite to halo mass ratios of 0.05, 0.03, and
    0.01 respectively.}
\label{fig:densevSD}
\end{figure*}

\begin{figure*}
  \centering
  \mbox{
    \mbox{\epsfxsize=3.0in\epsfbox{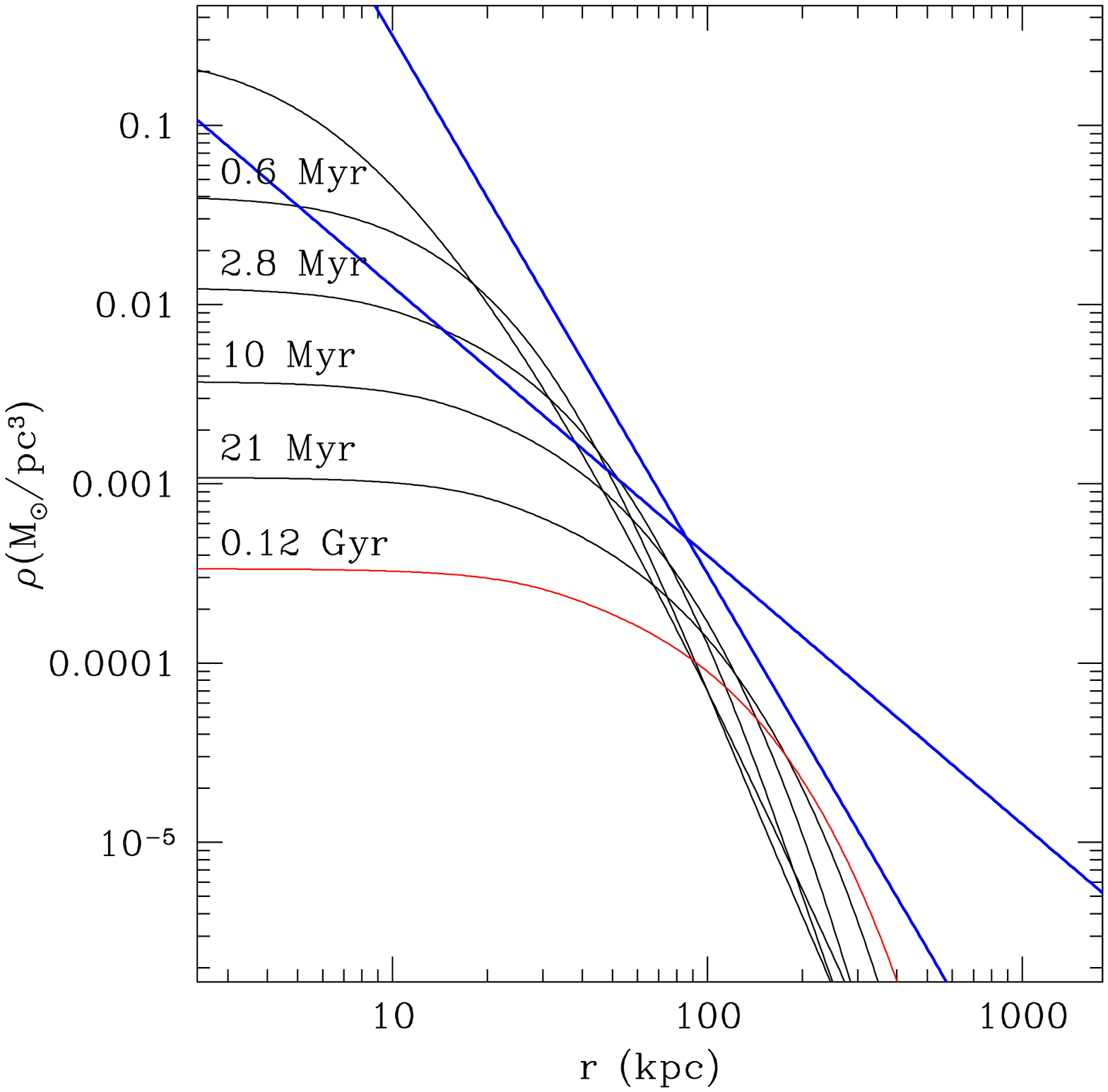}}
    \mbox{\epsfxsize=3.0in\epsfbox{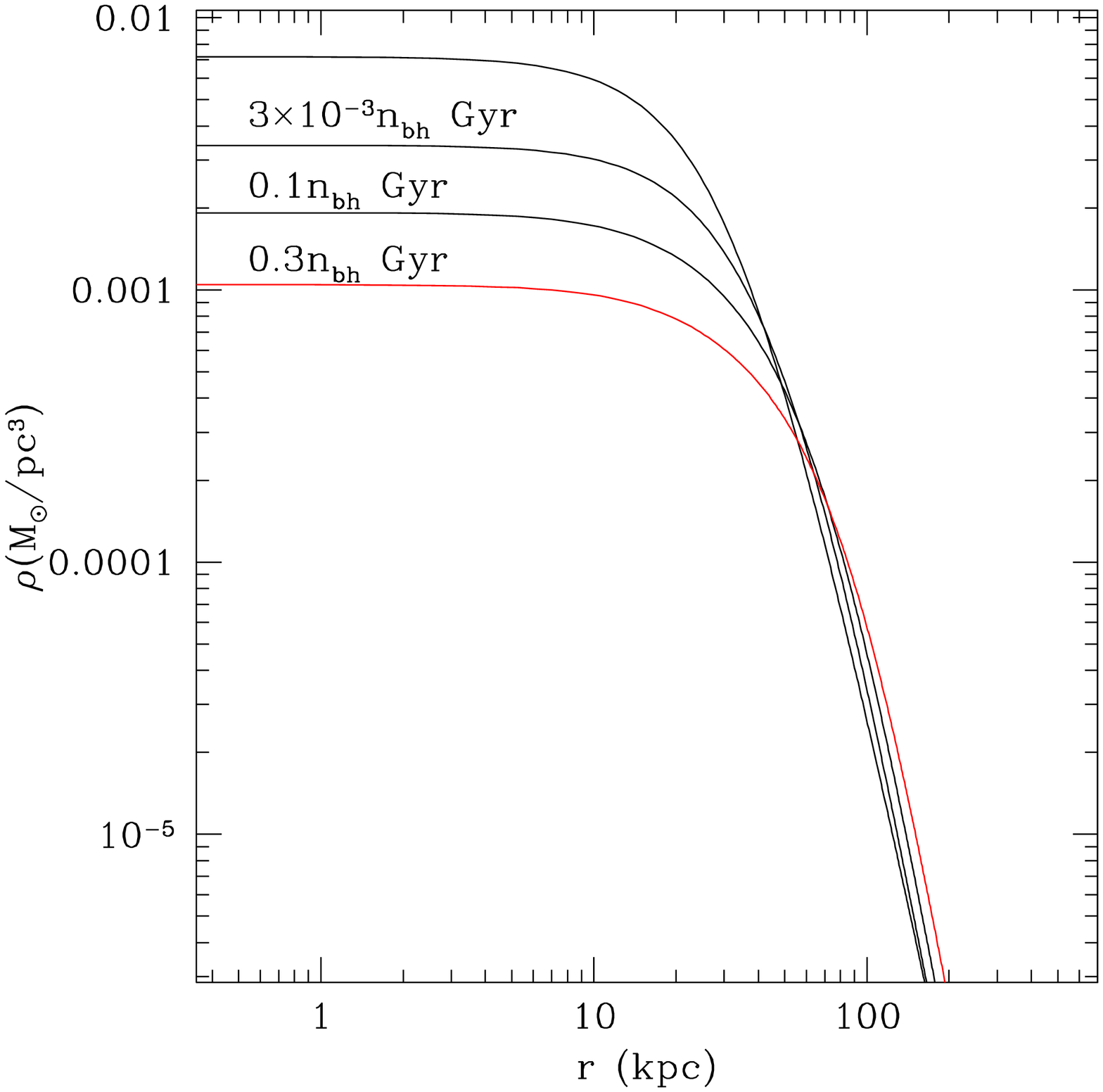}}
    }
  \caption{As in Fig. \protect{\ref{fig:densevol}}
    for a double power law model (eq. \protect{\ref{eq:NFWlike}}) with
    $\gamma=1$, $\beta=4$ and $\epsilon=0.1$.}
  \label{fig:densevP2}
  \caption{Evolution of a King $W_0=3$ profile under
    `black hole' noise.  The times for each curve are shown with the
    scaling for number of black holes per halo assuming that the black
    hole fraction is one ($f_{bh}=1$ in Table
    \protect{\ref{tab:scale}}).  Note that for the fiducial number,
    $n_{bh}=10^6$, the evolution time scale is uninterestingly large.}
  \label{fig:densevbh}
\end{figure*}

Evolution in the double power model (eq.\ref{eq:NFWlike}) shown in
Figure \ref{fig:densevP2} is similar to the cases shown in Figure
\ref{fig:densevol}.  Beginning with $\gamma=1$, $\beta=4$ and
$\epsilon=0.1$, the inner profile quickly attains the same shape as
those in Figure \ref{fig:densevol}, with a net transport of mass
outward and expansion.  The outer profile quickly approaches $r^{-3}$.
The evolution is an order of magnitude faster than the cases in Figure
\ref{fig:densevol} with the fiducial parameters given in Table 1
because the dynamical times in the cusp are smaller by the same
proportion.  Recall that the small time scales in Figure
\ref{fig:densevP2} are only consistent with the dynamics if there have
be many stochastic events.  For example, this better describes a
situation with $m_{enc}=0.01$ which increases the times in the figure
by 100.

Figure \ref{fig:densevbh} describes the evolution of a $W_0=3$ King
model due to point mass noise.  The evolution rate for this noise
source is uninteresting small unless the halo consists of fewer than
several hundred holes.  Note that because of the small amplitude,
curves in the figure are labeled for a single black hole instead of
the fiducial scaling given in Table \ref{tab:scale}.  There is little
point in characterizing this evolution in any more detail.

\section{Discussion}
\label{sec:discussion}

Two of our transient noise sources give double power law profiles: (1)
fly-by encounters; and (2) satellite mergers.  The evolved halo
profiles in each case are quite similar although the trajectories of
the perturbers are different.  This is explained by the excitation of
$l=m=1$ halo mode the low-frequency power in both cases.  The sloshing
mode results in transport of momentum that gives the shallower inner
power-law profile.  The response to the external $l=1$ multiple yields
the outer $r^{-3}$ profile.  The low-frequency power from the decaying
satellite comes from the speed of the decay, the changing orbital
frequencies in other words, not the orbital frequencies directly.  An
ensemble of non-decaying satellites drives evolution at harmonics
$l>2$ and is relatively weak.  For example, a halo of $10^6\msun$
black holes will cause only minor evolution during the 1 Gyr.

For a initially low-concentration haloes, the inner-profile becomes
steeper as it approaches a power law (cf. Fig. \ref{fig:densevol}ab).
For high-concentration haloes and all haloes after sufficient time, the
core radius slowly increases.  This is an artifact of the halo
expansion due to the energy added by the noise source.  The more
centrally concentrated the initial profile, the more rapidly the
asymptotic profile obtains.  In all cases, the double power law form
is evident after several hundred million years for fiducial parameters
(see Table \ref{tab:scale}).

More generally, these results suggest the evolved halo profile will
not depend on the noise source in general as long as there is power
near the low-order modes.  Lowered, truncated isothermal spheres (King
models) over an order of magnitude in concentration,
$r_{core}/r_{tidal}$, and Plummer spheres all converge to the double
power law form.  It is difficult to solve the collisional Boltzmann
equation (eqs.  \ref{eq:fptrans} and \ref{eq:fpeavg}) for initial
profiles with large dynamic range owing to numerical limitations.  To
address initially cuspy profiles, we softened the cusp (cf. eq.
\ref{eq:NFWlike}) and found that the same evolved double power law
form obtained and with much smaller time scale because the dynamical
times in the cusp are short.  For initial conditions without cores,
the $l=1$ response will be near the characteristic radius of the
initial profile.  For example, Weinberg (1998) shows that the $l=1$
mode peaks at the characteristic radius in a Hernquist profile.
Moreover, the evolution time scale for a softened cusp model is nearly
an order of magnitude shorter under the same noise spectrum as for
King models.  This further suggests that noise can drive the inner
profile of a cuspy dark matter halo to the asymptotic form quite
quickly.

These trends may help explain some of the recent trends in
cosmological simulations.  For isolated haloes, the most massive
substructure has decayed or disrupted within 1 Gyr.  Figures
\ref{fig:densevol} and \ref{fig:densevSD} show a well-defined inner
slope with exponent $\gamma=1.5$ by 1 Gyr.  This is consistent with
the recent work by Moore et al. (1998) and Klypin et al.  (2000).  For
longer evolution or noisier evolution, the inner exponent continues to
evolve, although slowly (cf. Fig. \ref{fig:densevol}). In particular,
this may help explain Jing \& Suto (2000) finding that the inner slope
depends on environment and the evolution of the scale radius $r_s$
(e.g. NFW).  However, as described in \S\ref{sec:models}, further
development needed to address the affect of cuspy initial profiles
will be necessary to predict this trend and will be the subject of a
later paper.

\section{Summary}
\label{sec:summary}

This paper describes the evolution of a halo due to transient noise
typical of the epoch of galaxy formation ($\tau\simless1\gyr$).  We
consider two transient noise sources: fly by encounters and merging
substructure (satellites).  Both result drive the halo toward double
power law profiles with inner exponent of approximately $-1.5$ and
outer power law exponent of $-3$, similar to the form proposed by NFW
and Moore et al.  Over a range of initial halo concentrations, the
double power laws obtain independent of noise source and initial
profile.

The dynamical mechanism is simply explained.  When one disturbs a halo
it rings.  This ringing is damped but the mostly weakly damped modes
shape the response and dominate the angular momentum transport
responsible for evolution.  Therefore, it doesn't matter how one hits
the halo, as long as the modes can be excited, the evolution looks the
same.  The dominant modes are low frequency and low harmonic order
(dipoles) and can be driven by a wide variety of transient noise
sources.  The outer power law exponent, $-3$, is due to the repetitive
excitation and response of the halo to the outer $l=1$ multipole.  The
common appearance of the $r^{-3}$ profile in n-body simulations
suggests a noise-driven origin.

This noise driven evolution provides a natural explanation for the
near universality of the halo profiles found in CDM simulations and
may provide an explanation for a spread of inner power law exponents.
Variation in the substructure at different scales (either due to the
CDM power spectrum or dynamic range of the simulation) and differences
in the initial halo profile will produce different exponents at the
same point in time.  Additional work will be required to make precise
predictions for these trends.  Nonetheless, this work shows that noise
naturally drives halo evolution with near-universal form.

\section*{Acknowledgements}
I thank Neal Katz for stimulating discussions and suggestions and
Enrico Vesperini for valuable comments on the manuscript.  This work
was support in part by by NSF AST-9529328.

\appendix

\section{Fokker-Planck evolution equation}
\label{sec:FP}

From the development in Paper 1, we have that equation
describing the evolution under noise processes is:
\begin{equation}
  {\partial f^\prime({\bf I}, t)\over\partial t} =
  \left[-{\partial\over\partial I_k^\prime}D_k^\prime +
    {\partial^2\over\partial I_k^\prime \partial I_l^\prime}
  D_{kl}^\prime\right] f^\prime({\bf I}, t),
\label{eq:fptrans}
\end{equation}
where the $D_k$ and $D_{kl}$ are the time derivative of the action
moments due to the noise process.  The isotropically averaged
Fokker-Planck becomes
\begin{eqnarray}
{\partial f(E)\over\partial\,t} &=& 
{1\over P(E)}{\partial\over\partial E}\left[
-\langle D_E\rangle_{iso} P(E) f(E) \right . \\ \nonumber
&& + \left. {\partial \over\partial
  E}\left(\langle D_{EE}\rangle_{iso} P(E) f(E)\right)\right]
\label{eq:fpeavg}
\end{eqnarray}
where $P(E)$ is the phase space volume between $E$ and $E+dE$.
The moment coefficients $D_{E}$ and $D_{EE}$ can not be strictly
interpreted as advection and diffusion.  In particular the
first-moment term may be positive or negative at different energies
depending on the structure of the response and its frequency spectrum.

\section{Numerical implementation}
\label{sec:numerical}

We solve the evolution equation (\ref{eq:fpeavg}) using a two-step
splitting technique following the work on globular cluster evolution.
First, the gravitational potential is held fixed and the phase-space
distribution function is evolved.  The Fokker-Planck equation is
finite-differenced in energy using the Crank-Nicholson scheme in
flux-conserving form.  Because the ``advection'' term can be both
positive and negative, the Chang-Cooper algorithm does not apply
(Chang \& Cooper 1970).  The energy variable
is remapped as described in Appendix \ref{sec:mapping} to improve the
dynamic range.  After exploring a number of possibilities, the range
of the energy grid extends from the centre to $E\rightarrow0$.  The
boundary conditions, then, are zero flux through the boundary.  In
practice, the outer boundary is set to $E=-\epsilon$ for some small
$\epsilon$ which corresponds to some large radius.  Second, the
potential and density distribution with new distribution is solved
iteratively to convergence.  To do this, the distribution of actions
is held fixed and a self-consistent spatial profile is determined by
computing a new density distribution from the phase-space distribution
function and the current gravitational potential.  One then computes a
new gravitational potential from the new density profile and
continues.  We construct the mapping between energy $E$ and angular
momentum $J$ and the radial actions by using the marching squares
algorithm on a two-dimensional table of radial action as a function
$E$, and angular momentum $J/J_{max}(E)$.  The profile recomputed when
central density has changed by 2\%.  Empirically, 40 or 100 mesh
points in the transformed energy grid are sufficient for low and high
concentration models, respectively.

The moment coefficients $D_{E}$ and $D_{EE}$ defined in Paper 1 are
expensive and recomputed only when central density has changed by 5\%.
The discrete sums in the action-angle series are typically truncated
at $l_{1\,max}=10$.  For low-order expansions here, $l\le4$, this is
sufficient to include at least 95\% of the total contribution.  We use
the biorthogonal basis described Weinberg (1999) and truncate the
expansion at $n_{max}=10$.  This basis uses direct solution of the
Sturm-Liouville equation to find a basis whose lowest order function
matches the equilibrium profile.  The results were checked in several
cases using both the Clutton-Brock (1973) and Hernquist \& Ostriker
(1992) bases along with tests of varying $l_{1\,max}$, $n_{max}$ and
energy grids with comparable results.  The asymptotic double power law
form reported here shows no signs of being a numerical artifact.

\section{Energy grid mapping}
\label{sec:mapping}

In order to to resolve the distribution function over a large dynamic
range, one can remap the energy to distribute mesh points more
uniformly between the cusp and outer profile.  The following mapping
as proved useful in globular cluster work (Chernoff \& Weinberg 1990)
and we will adopt it here.
\begin{equation}
  x = -\log\left({E\over(2-\alpha)E_0 - E}\right)
\end{equation}
where $E_0$ is the smallest (most bound) energy and $\alpha$ is a
fixed parameter.  For small values of $\alpha$, this transformation is
becomes purely logarithmic.  As $\alpha\rightarrow1$ the mapping
diverges as $E\rightarrow E_0$.  By carefully choosing values
$\alpha<1$, we can place more grid points in the centre to better
resolve the profile.  For these calculations, $\alpha=0.85$ is a good
choice.

Changing variables as described in Paper 1 yields:
\begin{equation}
  D_x = {\partial x(E)\over \partial I_j} D_j + 
  {\partial^2 x(E)\over \partial I_i\partial I_j} D_{ij}.
\end{equation}
where
\begin{eqnarray}
  {\partial x(E) \over \partial I_j} &=& {dx(E)\over dE}\Omega_j \\
  {\partial^2 x(E) \over \partial I_i \partial I_j} &=& 
  {dx(E)\over dE}{\partial\Omega_j\over\partial I_i} +
  {d^2x(E)\over dE^2}\Omega_i\Omega_j.
\end{eqnarray}
Altogether we have:
\begin{eqnarray}
  D_x &=& {dx(E)\over dE}\Omega_j D_j + 
  {dx(E)\over dE}{\partial\Omega_j\over\partial I_i}D_{ij} +
  {d^2x(E)\over dE^2}\Omega_i\Omega_j D_{ij} \nonumber \\ && \\
  \noalign{\leftline{and}}
  D_{xx} &=& \left({dx(E)\over dE}\right)^2 \Omega_i \Omega_j D_{ij}
\end{eqnarray}
In this mapped energy variable, the Fokker Planck equation
(cf. eq. \ref{eq:fpeavg}) becomes
\begin{eqnarray}
{\partial f(E(x))\over\partial\,t} &=& {1\over P(E(x))J_x(x)} \times
\nonumber \\
&& {\partial\over\partial x}\left[\rule{0pt}{15pt}
-D_x P(E(x))J_x f(E(x)) \right. \nonumber \\
&& + \left. {1\over J_x(x)} {\partial \over\partial
  x}\left(D_{xx} P(E(x)) J_x f(E(x))\right)\right] \nonumber \\
\label{eq:fpx}
\end{eqnarray}
where
\begin{equation}
  J_x(x) \equiv {dE\over dx} = -{ (2-\alpha)E_0e^{-x} \over (1 +
    e^{-x})^2}.
\end{equation}

\bsp

\label{lastpage}


\begin{thebibliography}{}

\bibitem[]{Alvares.etal:00}
Alvares, M., Shapiro, P.~R., and Martel, H. 2000,
\newblock {\em The Effect of Gasdynamics on the Structure of Dark Matter
  Halos},
\newblock {\tt astro-ph/0006203}.

\bibitem[]{Binney.Tremaine:87}
Binney, J. and Tremaine, S. 1987,
\newblock {\em Galactic Dynamics},
\newblock Princeton University Press, Princeton, New Jersey.

\bibitem[]{Chang.Cooper:70}
Chang, J.~S. and Cooper, G. 1970,
\newblock J. Comp. Phys., 6, 1.

\bibitem[]{Chernoff.Weinberg:90}
Chernoff, D.~F. and Weinberg, M.~D. 1990,
\newblock ApJ, 351, 121.

\bibitem[]{Clutton-Brock:73}
Clutton-Brock, M. 1973,
\newblock Astrophys. Space. Sci., 23, 55.

\bibitem[]{Frenk.etal:00}
{Frenk}, C.~S., {White}, S. D.~M., {Bode}, P., {Bond}, J.~R., {Bryan}, G.~L.,
  {Cen}, R., {Couchman}, H. M.~P., {Evrard}, A.~E., {Gnedin}, N., {Jenkins},
  A., {Khokhlov}, A.~M., {Klypin}, A., {Navarro}, J.~F., {Norman}, M.~L.,
  {Ostriker}, J.~P., {Owen}, J.~M., {Pearce}, F.~R., {Pen}, U.~., {Steinmetz},
  M., {Thomas}, P.~A., {Villumsen}, J.~V., {Wadsley}, J.~W., {Warren}, M.~S.,
  {Xu}, G., and {Yepes}, G. 1999,
\newblock \apj, 525, 554.

\bibitem[]{Hernquist.Ostriker:92}
Hernquist, L. and Ostriker, J.~P. 1992,
\newblock ApJ, 386, 375.

\bibitem[]{Jing.Suto:00}
{Jing}, Y.~P. and {Suto}, Y. 2000,
\newblock \apjl, 529, L69.

\bibitem[]{King:66}
King, I.~R. 1966,
\newblock AJ, 71, 64.

\bibitem[]{Klypin.etal:00}
{Klypin}, A., {Kravtsov}, A.~V., {Bullock}, J., and {Primack}, J. 2000,
\newblock in {\em submitted to ApJ, 16 pages, 10 figures, uses aastex and
  natbib.}, pp 6343.

\bibitem[]{Lacey.Ostriker:85}
Lacey, C.~G. and Ostriker, J.~P. 1985,
\newblock ApJ, 299, 633.

\bibitem[]{Moore.etal:99}
{Moore}, B., {Ghigna}, S., {Governato}, F., {Lake}, G., {Quinn}, T., {Stadel},
  J., and {Tozzi}, P. 1999,
\newblock \apjl, 524, L19.

\bibitem[]{Moore.etal:98}
{Moore}, B., {Governato}, F., {Quinn}, T., {Stadel}, J., and {Lake}, G. 1998,
\newblock \apjl, 499, L5.

\bibitem[]{NFW:97}
{Navarro}, J.~F., {Frenk}, C.~S., and {White}, S. D.~M. 1997,
\newblock \apj, 490, 493.

\bibitem[]{Spergel.Steinhardt:00}
Spergel, D.~N. and Steinhardt, P.~J. 2000,
\newblock Phys. Rev. Lett., 84, 3760.

\bibitem[]{Tittley.Couchman:99}
Tittley, E.~R. and Couchman, H. M.~P. 1999,
\newblock {\em Hierarchical clustering, the universal density profile, and the
  mass-temperature scaling law of galaxy clusters},
\newblock {\tt astro-ph/9911365}.

\bibitem[]{Tormen.etal:98}
{Tormen}, G., {Diaferio}, A., and {Syer}, D. 1998,
\newblock \mnras, 299, 728.

\bibitem[]{Vesperini.Weinberg:00}
{Vesperini}, E. and {Weinberg}, M.~D. 2000,
\newblock \apj, 534, 598.

\bibitem[]{Weinberg:98}
{Weinberg}, M.~D. 1998,
\newblock MNRAS, 299, 499.

\bibitem[]{Weinberg:99}
{Weinberg}, M.~D. 1999,
\newblock \aj, 117, 629.

\bibitem[]{Weinberg:00}
Weinberg, M.~D. 2000,
\newblock MNRAS, submitted (Paper 1).

\end{thebibliography}
\end{document}